\documentclass[twocolumn]{aastex631}
\usepackage{graphicx}
\usepackage{threeparttable}
\usepackage{amsmath}
\usepackage{bm}

\newcommand{\REV}[1]{\textcolor{black}{#1}}

\newcommand{\PMO}{\affiliation{Purple Mountain Observatory, Chinese Academy of Sciences, Nanjing 210023, China}}
\newcommand{\AJCTIAA}{\affiliation{Instituto de Astrof{\'i}sica de Andaluc{\'i}a (IAA-CSIC), Glorieta de la Astronom{\'i}a s/n, E-18008, Granada, Spain}}
\newcommand{\USTC}{\affiliation{School of Astronomy and Space Sciences, University of Science and Technology of China, Hefei 230026, China}}
\newcommand{\NJU}{\affiliation{School of Astronomy and Space Science, Nanjing University, Nanjing 210023, China}}
\newcommand{\AJCTNikolaevNU}{\affiliation{Petro Mohyla Black Sea National University, Mykolaiv 54000, Ukraine}} 
\newcommand{\AJCTNikolaevNA}{\affiliation{Mykolaiv Astronomical Observatory, Mykolaiv 54030, Ukraine}}
\newcommand{\AJCTCSIC}{\affiliation{Unidad Asociada al CSIC Departamento de Ingenier{\'i}a de Sistemas y Autom{\'a}tica, Escuela de Ingenier{\'i}as, Universidad de M{\'a}laga, C\/. Dr. Ortiz Ramos s/n, 29071 M{\'a}laga, Spain}}
\newcommand{\AJCTUMA}{\affiliation{Department of Algebra, Geometry and Topology, Facultad de Ciencias, Universidad de M\'alaga, Boulevard Luis Pasteur, s\/n, E-29071, M\'alaga, Spain}}
\newcommand{\AJCTUMASC}{\affiliation{Servicios Centrales de Investigaci{\'o}n, Universidad de M{\'a}laga, Boulevard Luis Pasteur, s/n, E-29071, M{\'a}laga, Spain}}

\newcommand{\AJCTYNAO}{\affiliation{Yunnan Observatories, Chinese Academy of Sciences, Kunming 650216, China}} 
\newcommand{\AJCTARIES}{\affiliation{Aryabhatta Research Institute of Observational Sciences (ARIES), Manora Peak, Nainital-263002, India}}
\newcommand{\AJCTUNAMIAE}{\affiliation{Instituto de Astronom\'{\i}a, Universidad Nacional Aut\'onoma de M\'exico,  Carr. Tijuana-Ensenada km.107, 22860 Ensenada, B.C., Mexico}} 
\newcommand{\AJCTUNAM}{\affiliation{Instituto de Astronomía, Universidad Nacional Aut{\'o}noma de M{\'e}xico, Apdo Postal 70-264, Cd. Universitaria, 04510 M{\'e}xico DF, M{\'e}xico}}
\newcommand{\AJCTSMA}{\affiliation{Sociedad Malague\~na de Astronom{\'i}a, Rep{\'u}blica Argentina 2, 29071 M{\'a}laga, Spain}}

\newcommand{\AJCTSKKU}{\affiliation{Department of Physics, Sungkyunkwan University, Seobu-ro 2066, Suwon, 16419 Korea}}
\newcommand{\AJCTUFS}{\affiliation{Department of Physics, University of the Free State, PO Box 339. Bloemfontein 930,  South Africa}}
\newcommand{\AJCTUCD}{\affiliation{School of Physics, O’Brien Centre for Science North, University College Dublin, Belfield, Dublin 4, Ireland}}
\newcommand{\AJCTNJUKLMAA}{\affiliation{Key Laboratory of Modern Astronomy and Astrophysics (Nanjing University), Ministry of Education, Nanjing 210093, China }}
\newcommand{\AJCTCSPDA}{\affiliation{San Pedro de Atacama Celestial Explorations (SPACE Celestial Explorations),San Pedro de Atacama, Chile}}
\newcommand{\AJCTSISP}{\affiliation{Swedish Institute of Space Physics (IRF), Bengt Hultqvists v{\"{a}}g 1, 981 92 Kiruna, Sweden}}
\newcommand{\UHFS}{\affiliation{University of Helsinki, Faculty of Science, Gustav Hällströmin katu 2, FI-00014, Finland}}
\newcommand{\UFUIPT}{\affiliation{Institute of Physics and Technology, Ural Federal University, Mira str. 19, 620002 Ekaterinburg}}

\newcommand{\LHGTITANS}{\affiliation{Millennium Nucleus on Transversal Research and Technology to Explore Supermassive Black Holes (TITANS)}}
\newcommand{\LHGMAS}{\affiliation{Millennium Institute of Astrophysics (MAS), Nuncio Monseñor Sótero Sanz 100, Providencia, Santiago, Chile}}

\newcommand{\LHGUV}{\affiliation{Instituto de F{\'i}sica y Astronom{\'i}a, Facultad de Ciencias,Universidad de Valpara{\'i}so, Gran Bretana 1111, Playa Ancha, Valpara{\'i}so, Chile}}

\newcommand{\INAFOASS}{\affiliation{INAF, Osservatorio di Astrofisica e Scienza dello Spazio di Bologna, via Piero Gobetti 93/3, 40024, Bologna, Italy}}
\newcommand{\INAFOAB}{\affiliation{INAF-Osservatorio Astronomico di Brera, Via E. Bianchi 46, 23807 Merate, LC, Italy}}
\newcommand{\INAFOAR}{\affiliation{INAF-Osservatorio Astronomico di Roma, Via di Frascati 33, 00040 Monte Porzio Catone, RM, Italy}}
\newcommand{\UMITA}{\affiliation{University of Messina, MIFT Department, Via F. S. D'Alcontres 31, Papardo, 98166 Messina, Italy}}
\newcommand{\USIDSAT}{\affiliation{Università degli Studi dell’Insubria, Dipartimento di Scienza e Alta Tecnologia, via Valleggio 11, I-22100 Como, Italy }}

\newcommand{\CASLab}{\affiliation{CAS Key Laboratory for Research in Galaxies and Cosmology, Department of Astronomy, University of Science and Technology of China, Hefei, 230026, China}}
\newcommand{\IDSS}{\affiliation{Institute of Deep Space Sciences, Deep Space Exploration Laboratory, Hefei 230026, China}}

\newcommand{\CMU}{\affiliation{McWilliams Center for Cosmology, Department of Physics, Carnegie Mellon University, 5000 Forbes Ave, Pittsburgh, 15213, PA, USA}}
\newcommand{\IOE}{\affiliation{Institute of Optics and Electronics, Chinese Academy of Sciences, Chengdu 610209, China}}
\newcommand{\NAOJ}{\affiliation{National Astronomical Observatory of Japan, National Institutes of Natural Sciences, Tokyo 181-8588, Japan}}
\newcommand{\LabPDE}{\affiliation{State Key Laboratory of Particle Detection and Electronics, University of Science and Technology of China, Hefei 230026, China}}
\newcommand{\NOAO}{\affiliation{National Optical Astronomy Observatory (NSF's National Optical-Infrared Astronomy Research Laboratory), 950 N Cherry Ave, Tucson Arizona 85726, USA}}
\newcommand{\TAMU}{\affiliation{Department of Physics and Astronomy, Texas A\&M University, College Station, TX 77843-4242, USA}}
\newcommand{\ITAMU}{\affiliation{George P. and Cynthia Woods Mitchell Institute for Fundamental Physics and Astronomy, Texas A\&M University, College Station, TX 77843-4242, USA}}

\shorttitle{GRB 240529A}
\shortauthors{Sun et al.}

\begin{document}
\author[0000-0003-1166-3814]{Tian-Rui~Sun}
\PMO

\author[0000-0001-9648-7295]{Jin-Jun~Geng}\thanks{E-mail: jjgeng@pmo.ac.cn}
\PMO

\author{Jing-Zhi Yan}\thanks{E-mail: jzyan@pmo.ac.cn}
\PMO

\author[0000-0002-7400-4608]{You-Dong Hu}\thanks{E-mail: huyoudong072@hotmail.com}
\INAFOAB
\AJCTIAA

\author[0000-0002-6299-1263]{Xue-Feng Wu}
\PMO
\USTC

\author[0000-0003-2999-3563]{Alberto J. Castro-Tirado}
\AJCTIAA
\AJCTCSIC

\author[0009-0006-5171-3723]{Chao~Yang}
\PMO

\author[0000-0002-7529-1792]{Yi-Ding~Ping}
\PMO

\author[0000-0002-5238-8997]{Chen-Ran Hu}  
\NJU

\author[0000-0001-7943-4685]{Fan Xu} 
\NJU

\author[0000-0001-7892-9790]{Hao-Xuan Gao}
\PMO

\author[0000-0002-9092-0593]{Ji-An Jiang}
\CASLab
\USTC
\NAOJ

\author[0009-0009-5585-4728]{Yan-Tian Zhu}
\USTC
\PMO

\author{Yongquan Xue}
\CASLab
\USTC

\author[0000-0002-7273-3671]{Ignacio P{\'e}rez-Garc{\'i}a}
\AJCTIAA

\author[0009-0004-7113-8258]{Si-Yu Wu}
\AJCTIAA

\author[0009-0009-4604-9639]{Emilio  Fern{\'a}ndez-Garc{\'\i}a}
\AJCTIAA

\author[0000-0001-7920-4564]{María D. Caballero-Garc{\'i}a}
\AJCTIAA

\author[0000-0002-7158-5099]{Rubén S{\'a}nchez-Ram{\'i}rez}
\AJCTIAA

\author[0000-0003-2628-6468]{Sergiy Guziy}
\AJCTIAA
\AJCTNikolaevNU
\AJCTNikolaevNA

\author{Ignacio Olivares}
\AJCTIAA

\author{Carlos Jesus P{\'e}rez del Pulgar}
\AJCTCSIC

\author{A. Castell{\'o}n}
\AJCTCSIC
\AJCTUMA

\author{Sebastián Castillo}
\AJCTUMASC

\author[0000-0002-6809-9575]{Ding-Rong Xiong}
\AJCTYNAO

\author{Shashi B. Pandey}
\AJCTARIES

\author[0000-0002-4711-7658]{David Hiriart}
\AJCTUNAMIAE

\author{Guillermo Garc{\'i}a-Segura}
\AJCTUNAMIAE
\AJCTIAA

\author[0000-0002-2467-5673]{William H. Lee}
\AJCTUNAM

\author{I. M. Carrasco-Garc{\'i}a}
\AJCTSMA

\author[0000-0001-6665-9631]{Il H. Park}
\AJCTSKKU
\author[0000-0001-8890-5418]{Petrus J. Meintjes}
\AJCTUFS 
\author[0009-0004-9747-7215]{Hendrik J. van Heerden}
\AJCTUFS 

\author[0000-0001-5108-0627]{Antonio Mart{\'i}n-Carrillo}
\AJCTUCD

\author[0000-0003-2931-3732]{Lorraine Hanlon}
\AJCTUCD
\author[0000-0003-4111-5958]{Bin-Bin Zhang}
\NJU
\AJCTNJUKLMAA

\author{Alain Maury}
\AJCTCSPDA

\author[0000-0002-8606-6961]{L. Hern{\'a}ndez-Garc{\'i}a}
\LHGTITANS
\LHGMAS
\LHGUV

\author[0000-0003-4268-6277]{Maria Gritsevich}
\AJCTSISP
\UHFS
\UFUIPT

\author[0000-0002-8860-6538]{Andrea Rossi}  
\INAFOASS

\author[0000-0003-2593-4355]{Elisabetta Maiorano}
\INAFOASS

\author[0000-0003-2910-6565]{Felice Cusano}  
\INAFOASS

\author{Paolo D'Avanzo}
\INAFOAB

\author{Matteo Ferro}
\USIDSAT
\INAFOAB

\author[0000-0002-2810-2143]{Andrea Melandri}
\INAFOAR

\author[0000-0002-4036-7419]{Massimiliano De Pasquale}
\UMITA

\author[0009-0000-0564-7733]{Riccardo Brivio}
\INAFOAB
\USIDSAT

\author[0000-0001-8060-1321]{Min Fang}
\PMO

\author[0000-0003-4200-4432]{Lu-Lu Fan}
\CASLab
\USTC
\IDSS

\author{Wei-Da Hu}
\TAMU
\ITAMU

\author{Zhen Wan}
\CASLab
\USTC

\author[0000-0001-7201-1938]{Lei Hu}
\CMU

\author{Ying-Xi Zuo}
\PMO

\author{Jin-Long Tang}
\IOE

\author[0000-0003-2100-3596]{Xiao-Ling Zhang}
\PMO

\author[0000-0003-3728-9912]{Xian-Zhong Zheng}
\PMO

\author[0000-0001-9327-0920]{Bin Li}
\PMO

\author[0000-0003-1297-6142]{Wen-Tao Luo}
\IDSS

\author[0000-0002-2237-3655]{Wei Liu}
\PMO

\author[0000-0003-1617-2002]{Jian Wang}
\LabPDE
\IDSS

\author[0000-0002-1463-9070]{Hong-Fei Zhang}
\LabPDE

\author{Hao Liu}
\LabPDE

\author{Jie Gao}
\LabPDE

\author{Ming Liang}
\NOAO

\author[0000-0002-4372-0759]{Hai-Ren Wang}
\PMO
\author{Da-Zhi Yao}
\PMO

\author{Jing-Quan Cheng}
\PMO

\author[0000-0002-1330-2329]{Wen Zhao}
\CASLab
\USTC

\author[0000-0002-7835-8585]{Zi-Gao Dai}
\USTC

\title{GRB 240529A: A Tale of Two Shocks}

\begin{abstract}
Thanks to the rapidly increasing time-domain facilities, we are entering a golden era of research on gamma-ray bursts (GRBs). 
In this {\it Letter}, we report our observations of GRB 240529A with the Burst Optical Observer and Transient Exploring System, the 1.5-meter telescope at Observatorio Sierra Nevada, the 2.5-meter Wide Field Survey Telescope of China, the Large Binocular Telescope, and the Telescopio Nazionale Galileo. 
The prompt emission of GRB 240529A shows two comparable energetic episodes separated by a quiescence time of roughly 400 s. 
Combining all available data on the GRB Coordinates Network, we reveal the simultaneous apparent X-ray plateau and optical re-brightening around $10^3-10^4$~s after the burst. Rather than the energy injection from the magnetar as widely invoked for similar GRBs, the multi-wavelength emissions could be better explained as two shocks launched from the central engine separately.
The optical peak time and our numerical modeling 
suggest that the initial bulk Lorentz factor of the later shock is roughly 50, which indicates that the later jet should be accretion-driven and have a higher mass loading than a typical one. The quiescence time between the two prompt emission episodes may be caused by the transition between different accretion states of a central magnetar or black hole, or the fall-back accretion process. A sample of similar bursts with multiple emission episodes in the prompt phase and sufficient follow-up could help to probe the underlying physics of GRB central engines. 
\end{abstract}

\section{Introduction}\label{sec:intro}
Gamma-ray bursts (GRBs) are the most powerful explosions in the Universe.
They are classified into the short/long GRBs according to their duration in gamma-rays, or the collapsar/merger origin GRBs depending on the progenitor that generated the explosion~\citep{Kumar15,Zhang18,AJCT2024}.
The prompt emission of GRBs is thought to result from the internal dissipation of the outflow launched from the central engine~\citep[e.g.,][]{Rees94}, while the subsequent multi-wavelength afterglows are the synchrotron radiation produced by the propagation of shocks in the circum-burst medium~\citep{Piran99,Sari99}.
X-ray observations by the Neil Gehrels {\it Swift} Observatory ({\it Swift}, \citealt{Gehrels04,Burrows05}) revealed diverse light curve after the burst, including features like steep decay, flares, plateau, and normal decay~\citep{Zhang06,Tang19}. Meanwhile, some GRBs exhibit re-brightenings in the optical bands~\citep[e.g.,][]{Kann24}.
These are characteristics that deviate from the expectations of the standard model, i.e., a simple outgoing blastwave that gives synchrotron radiations~\citep[e.g.,][]{Blandford76,Sari99,Huang00,WangXG2015}, therefore providing valuable opportunities to advance our understanding of GRBs' central engines.

In this Letter, we report our observations and analyses on GRB 240529A, a recently-discovered GRB with simultaneous apparent X-ray plateau and optical re-brightening. The rapid follow-ups by space-based and ground-based facilities globally result in a unique story in the GRB museum.  
This Letter is organized as follows: In Section 2, we describe the observations and data analysis process. Multi-wavelength modeling is present in Section 3, and relevant implications and constraints are discussed in Sections 4 and 5. We summarize our results in Section 6. 

\section{Observations} \label{sec:obs}

GRB 240529A is a long GRB, which first triggered the {\it Swift} Burst Alert Telescope (BAT, \citealt{Barthelmy2005BAT,GCNSwfit1}) at 2024-05-29 02:58:31 (UT), and was successively followed in the X-ray band by the X-Ray Telescope (XRT, \citealt{Burrows2004XRT,GCNSwfit1}) and the optical band by the Ultra-Violet/Optical Telescope (UVOT, \citealt{Roming2005UVOT,GCN_UVOT}). The refined BAT ground-calculated position is RA,Dec = 335.341$^{\circ}$, 51.557$^{\circ}$ (J2000) ~\citep{GCNSwfit2}.
The AstroSat satellite~\citep{AstroSat}, the Hard X-ray Modulation Telescope~\citep{HXMT}, and the Interplanetary Network~\citep{IPN} soon confirmed this detection.
It was found that this burst consists of two separated multi-peaked emission episodes by Konus-Wind~\citep{Konus-Wind}, among which the first episode starts at a gap time of $T_{\rm gap} \sim 400$~s earlier than the second one detected by {\it Swift}. These two comparable energetic burst episodes with a clear quiescence time seem rare in the literature. 

Multi-wavelength follow-ups have been carried out subsequently, ranging from X-rays~\citep{GCN_XRT} to the optical~\citep{GCN_GOTO,GCN_NOT,GCN_BOOTES,GCN_Skynet,GCN_WINTER,GCN_GROWTH,GCN_Konkoly,GCN_MAAO,GCN_SAO,GCN_Mondy,GCN_DFOT,GCN_AKO,GCN_OSN,GCN_MASTER} and radio bands~\citep{GCN_AMI-LA}. Its redshift is identified as $z = 2.695$ using OSIRIS+ at the 10.4 m GTC telescope at the Roque de los Muchachos Observatory~\citep{GCN_GTC}.
This gives an equivalent isotropic energy of $\simeq 10^{54}$~erg for each episode~\citep{Konus-Wind}.

\subsection{Ground Observations}
The Burst Optical Observer and Transient Exploring System (BOOTES, \citealt{ajct2023nature,huyoudong2023frontiers}) network observed the afterglow of GRB240529A with three telescopes for consecutive observations in clear filter:  BOOTES-6 (Dolores Pérez-Ramírez Telescope) 0.6m robotic telescope at Boyden Observatory, South Africa, BOOTES-5 (Javier Gorosabel Telescope) 0.6m telescope at Observatorio Astronómico Nacional en la Sierra de San Pedro Mártir, Mexico, and BOOTES-7 telescope at San Pedro de Atacama Space Observatory, Chile. The BOOTES-6 telescope observed the GRB 240529A starting at May. 29 2024, 03:05 UT (about 419~s after the BAT trigger).

We triggered the 1.5-meter telescope at Observatiorio Sierra Nevada (OSN, Granada, Spain) for the afterglow observations in $R$ and $Ic$ bands since Jun. 1 2024, at 01:06 UT ($\sim$ 2.9 days post burst). We also initiated a follow-up campaign with the 2.5 m Wide Field Survey Telescope (WFST) in the $g$ and $r$ bands separately after midnight of 29 May and on 30 May. 
WFST is a newly established photometric survey facility equipped with a 2.5 m diameter primary mirror installed near the summit of Saishiteng Mountain in northwestern China~\citep{2023WFST}.

Late-time observations of the field were performed with the Large Binocular Telescope (LBT) using the two twin LBC instruments equipped
with a UV filter $U_{spec}$ and the Sloan filters $g^\prime r^\prime i^\prime z^\prime$ on June 5, 2024~\citep{Rossi2024GCN.36655}, at the midtime 09:45 UT (7.28 days after the burst trigger). A second deep epoch was obtained in the $r^\prime i^\prime$ filters on June 12, 2024 at the midtime of 08:55 UT (14.25 days after the burst). Additionally, on the same night $J$-band imaging was also obtained (20:30 UT, $\sim$ 14.3 days after the trigger).

\subsection{Data Reduction}
The basic data reduction followed the standard CCDPROC procedure ~\citep{2015ascl.soft10007C}, and the astrometric solution was corrected with the Gaia-DR2 catalogue~\citep{gaiadr2ast,astrometrynet}. The astrometric solutions for WFST used a triangle match method~\citep{pingmakewcs} with the Gaia-DR2 catalogue. 

For the data from the BOOTES network, we registered all observation images into the same projection using the SWarp procedure ~\citep{swarp}. We performed a forced photometry method with the source extractor~\citep{bertinsext} with point-spread-function (PSF) results generated by PSFEx~\citep{psfex}. The observation of BOOTES networks was using the $clear$ filter, so we calibrated the observation results to $G$ magnitude in the Gaia-DR2 catalogue~\citep{gaiadr2} as magnitude references. 

\begin{figure*}
   \begin{center}
   \includegraphics[scale=0.5]{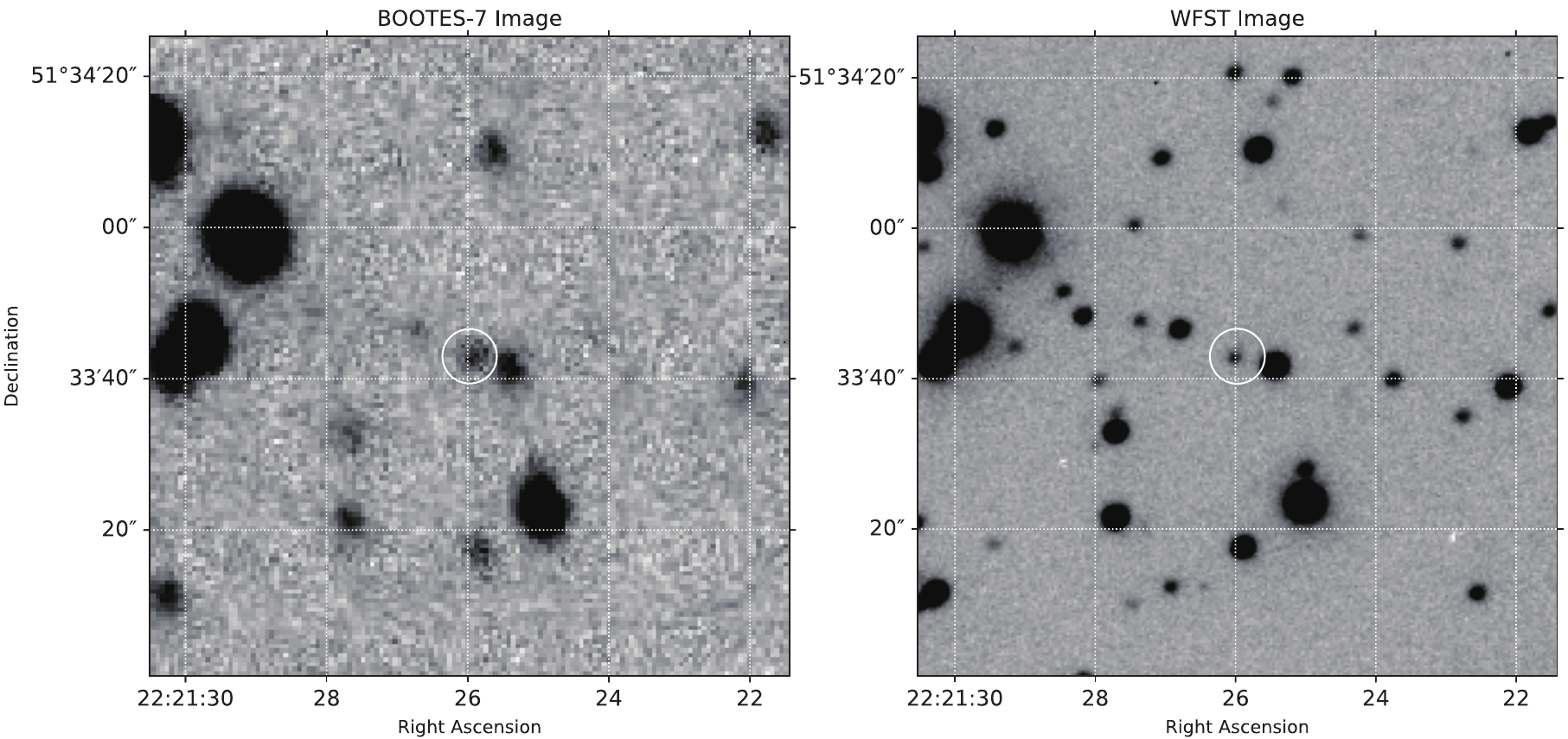}
   \caption{
   The left image exhibits the BOOTES-7 image observed at 09:44:40 on 29 May 2024 with a 60-second exposure and a magnitude limit of 18.3 mag in $clear$ filter. The right image exhibits the WFST image observed in the $g$ band at 18:29:01 on 29 May 2024 with a 180-second exposure and a magnitude limit of 22.1. Note that in both images, the circle indicates the position of GRB 240529A, with north registered as up.
   }
   \label{fig:obs}
   \end{center}
\end{figure*}

The photometry results for the WFST data were generated by a pipeline with the PSF photometry using the Source Extractor with the PSF results generated by PSFEx. The magnitude calibration of WFST observational data was calibrated against the Pan-STARRS DR1 catalogue~\citep{ps1catchambers16}. Representative images of GRB 240529A observed by BOOTES and WFST are shown in Figure~\ref{fig:obs}. 

The afterglow of GRB 240529A in the photometry images of the OSN telescope was very faint. We took images with the OSN telescope on Aug. 10 as the reference images for image subtractions. We registered all the images to the same projection with SWarp and performed image subtractions with the Saccadic Fast Fourier Transform package~\citep{sfft}. The photometry results were generated by forced photometry with photutils~\citep{photutils} on the difference images.
The observational results in the $R$ and $Ic$ bands for OSN observations were calibrated with the Pan-STARRS Data Release 1 catalogue and their bandpass transformations~\citep{ps1photo}.

All LBT data were reduced using the data reduction pipeline developed at INAF - Osservatorio Astronomico di Roma \citep{Fontana2014a} which includes bias subtraction and flat-fielding, bad pixel and cosmic ray masking, astrometric calibration, and coaddition.
The LBT telescope provided very deep images with magnitude limits reached 25.5 $mag$ in $i$ band from the coadded image. Since the point source of GRB 240529A is clearly detected in the LBT image, we performed PSF photometry using the standalone DAOPHOT package~\citep{daophot} on these coadded images. The observational results in the $g$, $r$, $i$ and $z$ bands for LBT observations were calibrated against the Pan-STARRS Data Release 1 catalogue. 
We cannot identify any emission from an extended source, i.e., host-galaxy, at the location of the afterglow. Indeed, in the late images we identify the possible host-galaxy to be offset (between 1 and 2 arcsec, $\approx$8-16 kpc) from the afterglow (RA (J2000) = 22$^{\rm h}$21$^{\rm m}$25$\fs$94, Dec. (J2000) = $+$51$^{\circ}$33\arcmin40\farcs6) from the LBT image, thus we will not consider the host contribution in our modelling of the light curve.

The TNG image was reduced using the jitter task from the ESO-eclipse package\footnote{\url{https://www.eso.org/sci/software/eclipse/}}~\citep{eclipse}. Photometric measurements were obtained through both aperture photometry and PSF-matched photometry with the DAOPHOT package within IRAF~\citep{iraf}, which were calibrated against nearby reference stars listed in the 2MASS catalogue~\citep{2mass}. 

All photometry results in a machine-readable table is available.

\section{Modeling} \label{sec:model}
\begin{figure*}
   \begin{center}
   \includegraphics[scale=0.9]{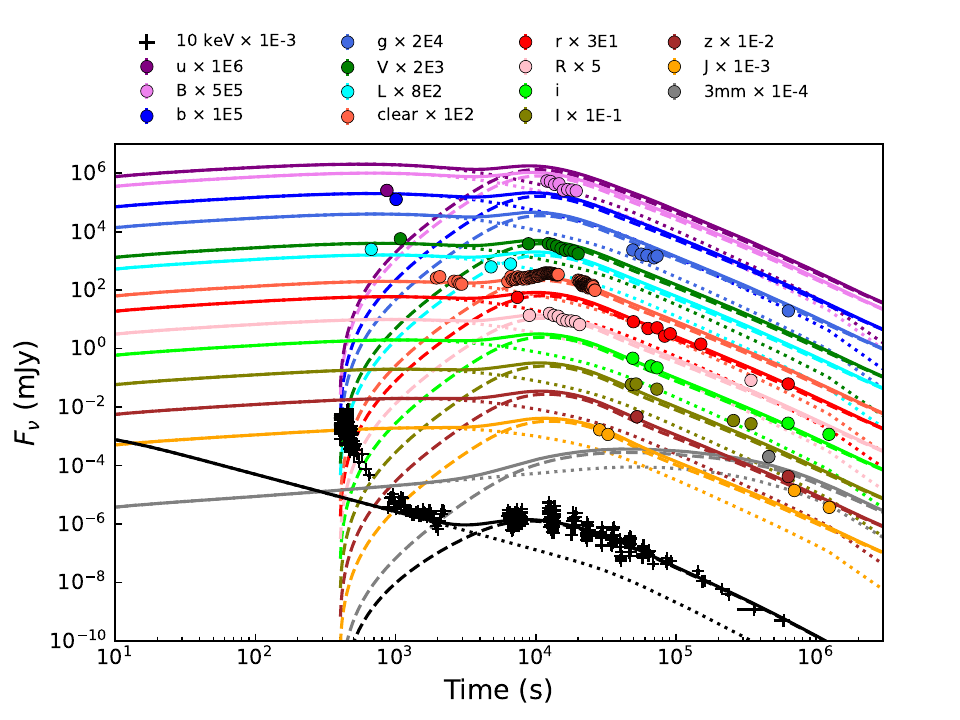}
   \caption{Modeling the multi-wavelength afterglow of GRB 240529A with the two external shocks. The dotted and dashed lines represent emissions from the former and the later shocks respectively, and the the solid lines are the sum of them.
   The X-ray data are taken from the {\it Swift}/XRT website (http://www.swift.ac.uk/xrt\_curves/01231488/). The optical data are taken from Section 2 and GCN circulars \citep{GCN_Skynet,GCN_WINTER,GCN_GTC,gcn36575_t193,GCN_GROWTH,GCN_Mondy,GCN_DFOT,gcn36654_mitsume,GCN_MAAO,GCN_AKO,GCN_Konkoly,gcn36734_jinshan,GCN_NOT,Rossi2024GCN.36655}. The radio data is taken from \cite{GCN_AMI-LA}.
   Note that the prompt emission and the steep decay of the later burst are not used in the fitting.
   The optical data have been corrected for Galactic extinction of $E(B-V) = 0.29$~mag~\citep{Schlafly11}.
   The excess of the modeling flux at the short wavelengths hints at the potential extinction of the host galaxy.}
   \label{fig:fit}
   \end{center}
\end{figure*}

Combining all the available data, the multi-wavelength afterglow emissions of GRB 240529A are exhibited in Figure~\ref{fig:fit}.
During the prompt phase, the released isotropic gamma-ray energy of each episode is $\simeq 10^{54}$~erg~\citep{Konus-Wind}.
Rather than a power-law decay in normal bursts, optical afterglows show re-brightenings with a peak time of $\simeq 1.2 \times 10^4$~s in GRB 240529A, accompanied by an apparent X-ray plateau.  
The long-term monitoring of XRT indicates that the temporal index changes from $-1.76 \pm 0.03$ to $-2.74 \pm 0.21$ at $\sim 10^5$~s, which may be a result of the jet beaming break~\citep{Piran00}. \REV{However, no significant simultaneous break is present in optical bands due to insufficient observation sampling around this time.} 
In the following, two possible scenarios are explored to understand the characteristics of GRB 240529A.

\subsection{The Magnetar Scenario}
Energy injection processes~\citep{Dai98a,Dai98b,Fan06,Geng13,vanEerten14} or the refreshing of the forward shock (FS) by late-arriving shells~\citep{Zhang01} are often invoked to explain the production of an X-ray plateau. 
On the other hand, it has been suggested that the central engines of some GRBs are newly born magnetars, from which a wind of Poynting-flux~\citep{Dai98b,Zhang06} or electron-positron pairs could inject into the FS and result in an X-ray plateau~\citep{Dai04,Yu07a,Yu07b}. 
This magnetar model has successfully been applied in explaining the simultaneous X-ray plateau and the optical re-brightening in some GRB afterglows~\citep{Geng16,Geng18a}. However, we realize that this scenario requires an unusually-high stellar spin in the case of GRB 240529A, as estimated in the following paragraph, unless a Poynting-flux-dominated wind or an electron-positron-pair-dominated wind that is available for energy injection is beamed in a jet.

The wind luminosity from the rotating magnetosphere of the magnetar is~\citep[e.g.,][]{Shapiro83,Xu01,Contopoulos06}
\begin{equation}
L_{\rm w} = 9.6 \times 10^{46} B_{\rm NS,14}^{2} R_{\rm NS,6}^{6} P_{0,-3}^{-4} \left(1 + \frac{t_{\rm obs}}{T_{\rm sd}} \right)^{-2} \mathrm{erg~s}^{-1},    
\end{equation}
where $B_{\rm NS}$ is the strength of the surface magnetic field of the
magnetar, $R_{\rm NS}$ is the radius, $P_{0}$ is the initial spin period, $T_{\rm sd} \simeq 2 \times 10^5 (1+z) I_{45} B_{\rm NS,14}^{-2} R_{\rm NS,6}^{-6} P_{0,-3}^{2}~\mathrm{s}$ is the spin-down timescale of the magnetar, and $I$ is its moment of inertia.
The convention $Q_x = Q/10^x$ in cgs units is adopted hereafter.
If we attributed the X-ray luminosity near the end of the plateau to the dissipation of the magnetar wind with an efficiency of $\eta$, i.e., $L_{\rm X} \simeq \eta L_{\rm w} (T_{\rm sd})$, one can obtain
\begin{eqnarray}
P_0 &\leq& 0.1 \left(\frac{\eta}{0.1} \right)^{1/2} \left(\frac{I}{10^{45} ~\mathrm{g~cm^2}} \right)^{1/2} \\
& &\times \left(\frac{L_{\rm X}}{10^{49}~\mathrm{erg~s}^{-1}} \right)^{-1/2} \left(\frac{T_{\rm sd}}{10^4~\mathrm{s}} \right)^{-1/2}~\mathrm{ms}, \nonumber
\end{eqnarray}
which requires a newborn sub-millisecond magnetar for GRB 240529A with $L_{\rm X} = 4 \pi D_{\rm L}^2 F_{\rm X} / (1+z)^{1-\beta_{\rm X}} \simeq 2 \times 10^{49}$~erg~s$^{-1}$ and $T_{\rm sd} \simeq 10^4$~s as inferred from the X-ray data. Here, an observed flux of $F_{\rm X} \simeq 3 \times 10^{-10}$~erg~cm$^{-2}$~s$^{-1}$ and a spectral index of $\beta_{\rm X} \simeq 1.2$ by the XRT is used.
Such a rapidly rotating neutron star is outside of the magnetars associated with GRBs with plateaus~\citep{Stratta18} and beyond the prediction of any equation of state for dense nuclear matter. However, it could be a strange quark star~\citep{Frieman89}. This is because the gravitational-radiation-driven r-mode instability of a rapidly-spinning star is highly suppressed due to a large bulk viscosity associated with the non-leptonic weak interaction among quarks and thus a newborn strange quark star can rotate at a sub-millisecond period~\citep{Dai16}.

\subsection{The Two-Shock Scenario}
Considering that there are two energetic emission episodes in the prompt phase, it is reasonable to suppose that each episode would drive an FS into 
the circum-burst environment and the observed flux is the superposition of the emission from these two shocks.  
Energy extraction processes directly from the central engine do not suffer from the energy budget issue discussed in Section~3.1. 
In this scenario, the early observed decaying optical afterglow is contributed by the former shock, while the optical re-brightening is due to the onset of the emission from the later shock.
Since the emission of the former shock is obscured by that of the later shock for $t_{\rm obs} \ge 5 \times 10^3$~s, it is difficult to sufficiently constrain relevant parameters of the former shock. Hence we mainly focus on the properties of the later shock below. Assuming that the initial Lorentz factor and the isotropic kinetic energy of the later shock are $\Gamma_{0}$ and $E_{\mathrm{K,iso}}$,
the peak time of the optical re-brightening corresponds to the end of the coasting phase of the blastwave in the standard GRB shock model~\citep{Blandford76,Sari99}, i.e.,  
\begin{equation}
t_{\rm peak} \simeq 10^4 \left(\frac{E_{\mathrm{K,iso}}}{10^{54}~\mathrm{erg}} \right)^{1/3} \left(\frac{n}{1~\mathrm{cm}^{-3}} \right)^{-1/3} \left( \frac{\Gamma_0}{50}\right)^{-8/3}~\mathrm{s},
\end{equation}
where $n$ is the number density of the ambient medium experienced by the later shock. The relatively long peak time of $\simeq 10^4$~s of GRB 240529A indicates that $\Gamma_0$ of the later shock should be significantly smaller than the typical value of $\sim 300$~\citep{Ghirlanda18}.

\begin{figure*}
   \begin{center}
   \includegraphics[scale=0.4]{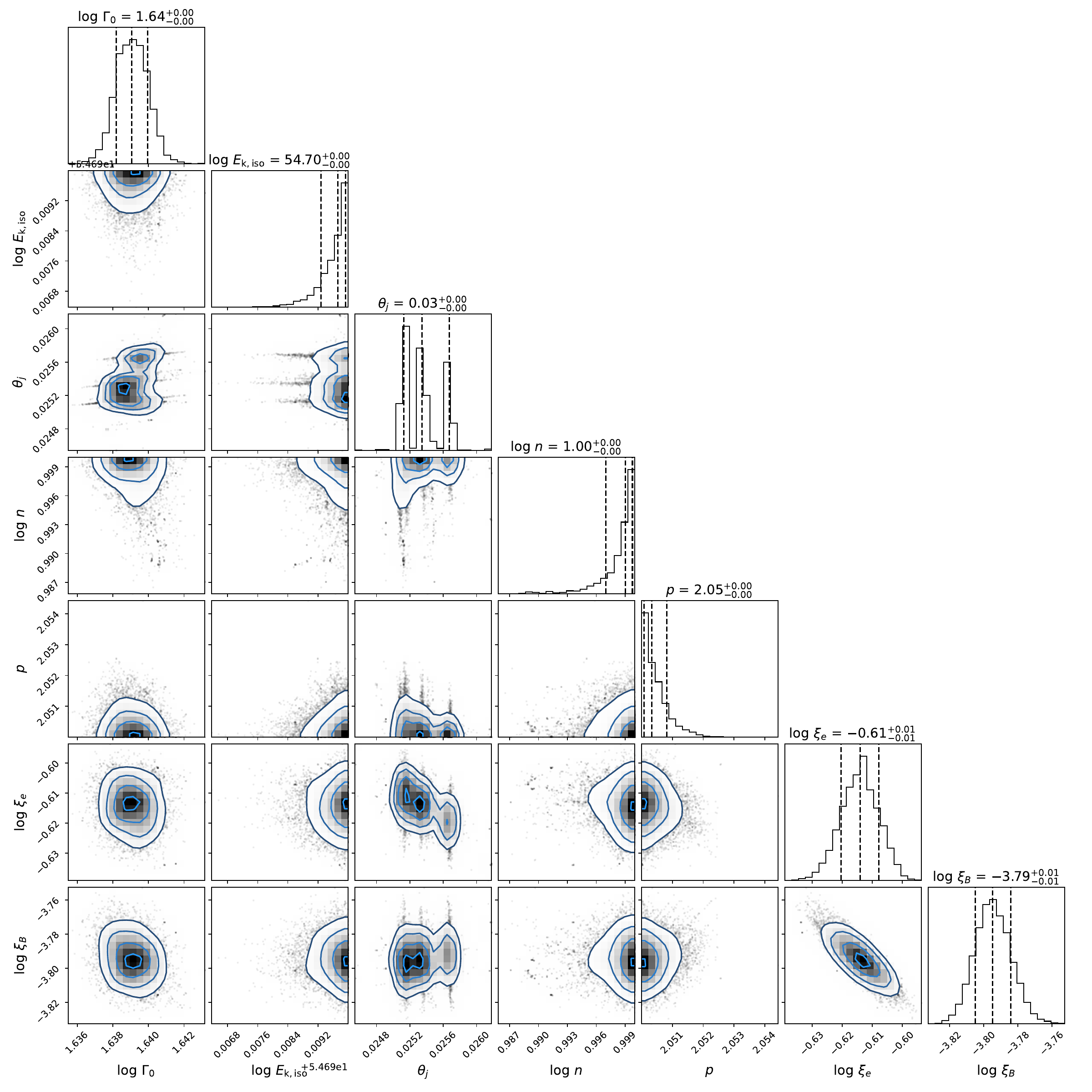}
   \caption{The corner plot of the fitting result in Figure~\ref{fig:fit}.
   It shows all the one and two dimensional projections of the posterior probability distributions of seven model parameters listed in Table~\ref{tab:para}. The 1-dimensional histograms are marginal posterior distributions of these parameters, in which the vertical dashed lines indicate the 16th, 50th, and 84th percentiles of the samples, respectively.
   }
   \label{fig:corner}
   \end{center}
\end{figure*}

Detailed numerical calculations to fit the multi-wavelength data are performed to demonstrate the consistency of this scenario.
The generical dynamical equations~\citep{Huang99,Huang00,Peer12} incorporating the jet sideways expansion~\citep{Granot12} are adopted to describe the jet dynamics.
We then solve the continuity equation of electrons accelerated by the FS in the energy space, from which multi-wavelength afterglows are derived~\citep{Geng18b,Gao24}.
The ``thermal'' electrons due to the inefficient shock heating are included in our calculations~\citep{Giannios09,Gao24}.
Specifically, the ratio of the non-thermal electron energy to the whole shocked electron energy $\delta$ is taken as $0.8$.
Seven parameters, i.e., $E_{\rm K,iso}$, $\Gamma_0$, the half-opening angle of the jet $\theta_{\rm j}$, the accelerated electron spectral index $p$, the equipartition parameter for shocked electrons $\xi_{{\rm e}}$ and magnetic fields $\xi_{B}$, and medium density number $n$ are left free for each shock.
Parameters of the former shock are chosen after several trials by hand. In contrast, parameters of the later shock are constrained by embedding our numerical module into the prevailing Markov chain Monte Carlo Ensemble sampler called emcee~\citep{Foreman-Mackey13}, as in our previous works~\citep{Geng16,Xu22}.  
The posteriors are listed in Table~\ref{tab:para} and Figure~\ref{fig:corner},
from which the derived initial bulk Lorentz factor of $\Gamma_0 = 43.6^{+0.1}_{-0.1}$ is consistent with the analytical analysis above.
Since the highest photon energy of $\simeq$ 3 MeV announced by Konus-Wind~\citep{Konus-Wind} is less than the critical value of $\Gamma_0 m_{\rm e} c^2 /(1 + z) \simeq 6$~MeV, the photons would not suffer from the annihilation during the prompt phase. Detailed spectral analyses may give a more robust low limit for allowed values of $\Gamma_0$.
Due to the degeneracy between $E_{\rm K,iso}$ and $n$, and the uncertainty of radiation efficiency during the prompt phase, $E_{\rm K,iso}$ always tends to be a large value, i.e., the upper boundary of the priors.
\REV{The parameters of the former shock are chosen to interpret the early decaying optical afterglow within $\sim 3 \times 10^3$~s (i.e., the coasting phase is much short) and to ensure that the X-ray afterglow does not exceed the observed flux of the second prompt episode. The scarcity of data and the strong degeneracy among parameters render these parameter values solely for reference purposes. As the afterglow temporal indices of one shock usually get steeper with time, it is reasonable to assume that the emission of the former shock is much fainter than that of the later shock. Therefore, how the early emission is modeled has little influence on fitting to the rest of the data.}
Note that the ambient number density for the two shocks launched subsequently is not necessarily the same since the leading one would modify the environment. 

\REV{The observational flux density of the J band exceeds our afterglow modeling at the late time around $10^6$~s. This exceeding flux corresponds to a component with a luminosity level of $10^{44}$~erg~s$^{-1}$ in the ultraviolet band at the rest frame, which could be emissions from the shock breakout in a supernova~\citep{Gezari15}, a potential super-luminous supernova~\citep{Drout14}, or other underlying sources. Its exact origin is highly uncertain and out of the scope of this article.}

\section{Discussion}
As the two-shock scenario could well explain the multi-wavelength emission of the GRB 240529A, the underlying physics to launch two energetic jets separately from the central engine remains puzzling. Three possible origins are discussed here. 

First, if the central engine is a magnetar, the bursting phase and the quiescence time may be interpreted as the transition between the accretion and propeller phases, regulated by the ordering of the magnetospheric radius and the corotation radius~\citep{Dai12,DallOsso23}.
The magnetospheric radius (a fraction of the Alfvèn radius) is $r_{\rm m} \propto \mu^{4/7} \dot{m}^{-2/7}$ and the corotation radius writes as $r_{\rm co} = (G M P^2 / 4 \pi^2)^{1/3}$, where $\mu$ is the magnetic
moment and $\dot{m}$ is the mass accretion rate.
\REV{The infant magnetar may take time to amplify its surface magnetic field and establish a relatively stable configuration~\citep{Raynaud20}, during which the accretion would occur and produce the first emission episode. When the magnetic field becomes strong enough, the emission ends as the system enters into the propeller regime ($r_{\rm m} \ge r_{\rm co}$). After a period of inactivity, the accretion will start again once the magnetar spin slows down to get into the accretion regime ($r_{\rm m} \simeq r_{\rm co}$), and hence produce the second emission episode.}
Assuming the burst luminosity to be $\propto \dot{m} c^2$, then the quiescence time between two burst episodes gives 
\begin{equation}
T_{\rm sd} \le \frac{T_{\rm gap}}{(\bar{L}_1/\bar{L}_2)^{6/7}-1} \equiv Y,    
\end{equation}
where $\bar{L}_1$ and $\bar{L}_2$ are the average luminosity of the former and later bursts, respectively.
Taking $\bar{L}_1/\bar{L}_2 \simeq 3$ and $T_{\rm gap} \simeq 400$~s, 
it further constrains the surface magnetic field strength to be
\begin{eqnarray}
B_{\rm NS} &\ge& 6 \times 10^{15} \left(\frac{Y}{130} \right)^{-1/2} 
\left(\frac{I}{10^{45}~\mathrm{g~cm^2}} \right)^{1/2} \\ 
& &
\times \left(\frac{P_0}{1~\mathrm{ms}} \right) \left(\frac{R_{\rm NS}}{10^6~\mathrm{cm}} \right)^{-3}~\mathrm{G},   \nonumber
\end{eqnarray}
which is consistent with the magnetar assumption.

Second, if the central engine is a rotating black hole (BH), it is suggested that
relativistic jets could be launched from the magnetically arrested disk~\citep{Tchekhovskoy11}. In this scenario, the quiescence time longer than $10$~s may correspond to the time in which the disk undergoes the interchange instability and the flux diffuses out into the disk~\citep{Lloyd-Ronning16}. The activity may restart after a viscosity timescale of $\tau_{\rm vis}$.
Under a simple Shakura–Sunyaev prescription of the disk around a BH with a mass of $M_{\rm BH}$, we have~\citep[e.g.,][]{Lloyd-Ronning16,Lei17,Liu17}
\begin{equation}
\tau_{\rm vis} \simeq 1 \left(\frac{\alpha}{0.1} \right)^{-1} \left(\frac{H/R}{0.3}\right)^{-2}  \left(\frac{R}{30~R_{\rm g}} \right)^{3/2} \left(\frac{M_{\rm BH}}{5~M_{\odot}} \right)~\mathrm{s},
\end{equation}
where $\alpha$ is the viscosity parameter, $H/R$ is the ratio of disk height $H$ to radius $R$, $R_{\rm g}$ is the Schwarzschild radius and $M_{\odot}$ is the solar mass. 
The relatively longer $T_{\rm gap}$ here may indicate that the $H/R$ is somehow smaller. Also, other reasons like fall-back debris~\citep{CannizzoLee90,Cannizzo09,LeeRamirez2009} can be invoked to explain for the second energetic burst.

At last, the central engine may be a rapidly rotating supramassive neutron star initially that collapses into a black hole later~\citep{VietriStella1998,Lasky14,Ravi14,Falcke14}.
The collapse time ($\le T_{\rm gap}$) is in the range of $\sim 10 - 4.4 \times 10^4$~s~\citep{Ravi14}. In this case, since the disk formation is seemingly unfeasible after the collapse of an isolated neutron star~\citep{Margalit15}, additional fall-back materials from the progenitor are still required to produce the second burst. 

\REV{All these possible mechanisms for the long-standing period between the two prompt episodes are principally consistent with the two-shock scenario for afterglow modeling. 
The significantly smaller bulk Lorentz factor for the second episode suggests that the baryon loading of the second episode is 
larger than that of a typical burst, supporting an accretion-driven origin. However, it is hard to tell whether the central engine is a magnetar or a black hole without clues on the progenitor star type or other robust evidence directly from the central compact star, a critical issue among most GRBs. On the contrary, the apparent X-ray plateau of GRB 240529A reminds us to be careful when we immediately associate the plateau feature with a central magnetar. Had Konus-Wind not detected the former emission episode, such misleading would have occurred.}

A single jet could be unstable to kink or sausage instabilities caused by toroidal magnetic fields and produce several luminous knots~\citep[e.g.,][]{Meyer13,Bromberg16}. These knots may be responsible for different emission episodes at the prompt stage. However, it is not easy for these knots to drive relativistic GRB shock and produce subsequent GRB afterglow emission episodes.

\begin{table*}
\centering
\caption{Parameters used in the modeling of the afterglow of GRB 240529A.\label{tab:para}}
\begin{threeparttable}
\begin{tabular}{lcc}
\toprule Parameters & Former Shock~\tnote{a} &  Later Shock \\
$E_{\rm K,iso}$~($10^{54}$~erg) &  1.0    &  $5.0^{-0.001}_{+0.001}$ \\
$\Gamma_0$                      &  800    &  $43.6_{-0.1}^{+0.1}$    \\
$\theta_{\rm j}$ (rad)          & 0.05    &  $0.03_{-0.001}^{+0.001}$ \\
$p$                             &  2.4    &  $2.05_{-0.001}^{+0.001}$   \\
$\xi_{{\rm e}}$                 &  0.1    &  $0.24_{-0.004}^{+0.003}$   \\
$\xi_{B}$                       &  $2.0 \times 10^{-4}$   &  $1.6_{-0.04}^{+0.04}$ $\times 10^{-4}$  \\
$n$~(cm$^{-3}$)                 &  7.0    &  $10.0_{-0.05}^{+0.02}$ \\
\hline
\end{tabular}
\begin{tablenotes}
\item[a] It is hard to constrain the parameters of the former shock strictly. 
\end{tablenotes}
\end{threeparttable}
\end{table*}

\section{Model Applicability and Constraints}
\REV{
We now discuss the application scope of the two-shock scenario
in the face of the diversity of GRB prompt emission and afterglows.
There are GRBs with two equivalent well-separated peaks in their prompt emission with no rebrightening in their optical afterglow. 
This could be due to several reasons.
First, the catch-up process of the two shocks would merge them into one blastwave soon if the separation time between two peaks is relatively short and the later shock is moving faster. 
Moreover, shock parameters like kinetic energy of the later shock may produce the afterglow emission fainter than that of the early shock.
In addition, the superposition of the two emission components with either fast or slow decaying behavior may not necessarily lead to the rebrightening, but a smooth temporal transition only.
Hence it is not easy to identify the afterglow emission of the later shock with confidence from the observations like GRB 240529A.}

\REV{
For GRBs with one main emission episode in the prompt phase and rebrightenings in their optical afterglow, the traditional energy injection scenarios should be preferred only if they do not suffer from the energy budget issue of the central engine (Section 3.1).}


\section{Conclusions} \label{sec:conclusion}
In this {\it Letter}, we report our observations of the recently detected GRB 240529A. 
Our data show clear optical re-brightening with a peak time of $\simeq 1.2 \times 10^4$~s.
The information on prompt emission motivates us to interpret the apparent X-ray plateau and the optical re-brightening of GRB 240529A using the two-shock scenario, rather than the traditional magnetar-wind-injection scenario. The peak time of the optical and X-ray afterglow suggests a relatively smaller initial bulk Lorentz factor of $\Gamma_0 \sim 50$ for the later shock,
which is also confirmed by the multi-wavelength afterglow modeling using our numerical method~\citep{Geng18b,Gao24}.

The prompt energy and $\Gamma_0$ of the second episode do not follow the relationship revealed by other bursts with much earlier peak time~\citep{Liang10,Yi17}. This indicates that the baryon loading of the second episode should be larger than that of typical bursts. This high baryon loading may prefer a neutrino-driven jet from a hyper-accreting black hole rather than a magnetically dominated jet~\citep{Lei13}. In comparison with the first episode, which could be a cleaner jet by extracting the rotation energy of a central black hole~\citep{Blandford77}, an accretion-driven jet for the second episode could be reasonable if the spin of the black hole or the surrounding magnetic field strength decreases significantly after a period of 400 s.

In some GRBs, there is a weak precursor before the main burst with the time gap ranging from $\sim 1$~s to $10^3$~s~\citep{Burlon08}. Bursts composed of two comparable main phases like GRB 240529A are unusual.
It may be due to the difficulty of identifying possible candidates from bursts with multiple peaks and episodes of different energy intensities in practice, or the rarity of the physical mechanisms themselves. 

Shock parameters like $E_{\rm K,iso}$ and $n$ are not well constrained in our fit due to the degeneracy between parameters and the radiation efficiency that is poorly known. Nevertheless, current fitting results show that the radiation efficiency of the first episode is higher than that of the second episode, which is consistent with the change of jet property by different energy extraction mechanisms mentioned above, i.e., from Poynting-flux dominated~\citep{Thompson94,Zhang11} to baryon-matter dominated~\citep{Popham99,Narayan01,Gu06,Lei13}.
As more adequate multi-wavelength data from the GeV to the radio band are crucial to narrowing down the parameter space, dedicated follow-up schemes of GRB afterglow are eagerly invoked to advance our understanding of GRBs in the current era.

Our work addresses the significance of analyses in the synergy of the prompt emission and the afterglow emission, which will also be shown in a series of works soon. The two active episodes separated by a quiescence time of $\sim 400$~s may be caused by the transition between different accretion states of a central magnetar or black hole, or the fall-back accretion process. High-cadence multi-wavelength monitoring of more similar bursts may shed light on the physics of the central engine and progenitors. 
 

\vspace{\baselineskip}
{
We thank Mao-Kai Hu for useful discussions.
This study is partially supported by the National Natural Science Foundation of China (grant Nos. 12273113, 12321003, 12393812, 12393813, and 12025303),
the National SKA Program of China (grant Nos. 2020SKA0120302 and 2022SKA0130100), the International Partnership Program of Chinese Academy of Sciences for Grand Challenges (114332KYSB20210018). JJG acknowledges support from the Youth Innovation Promotion Association (2023331). 
AJCT acknowledges support from the Spanish Ministry projects PID2020-118491GB-I00 and PID2023-151905OB-I00 and Junta de Andalucia grant P20\_010168 and from the Severo Ochoa grant CEX2021-001131-S funded by MCIN/AEI/ 10.13039/501100011033.
A. Rossi and E. Maiorano acknowledge support by PRIN-MIUR 2017 (grant 20179ZF5KS).
F.Cusano acknowledges support from the INAF project Premiale Supporto Arizona \& Italia.

MG acknowledges the Academy of Finland project No. 325806.
The program of development within Priority-2030 is acknowledged for supporting the research at UrFU (04.89).
LHG acknowledges funding from ANID programs: FONDECYT Iniciación 11241477, and Millennium Science Initiative Programs NCN$2023\_002$ and ICN12\_009. LH acknowledges support from Science Foundation Ireland grant 
19/FFP/6777.  GGS acknowledges support from DGAPA-UNAM PASPA grant. 
YDH acknowledges financial support from INAF through the GRAWITA Large Program Grant (ID: 1.05.12.01.04). I.H.P. acknowledges the support from NRF-2017K1A4A3015188 and NRF-2021R1A2B5B03002645.
TRS and HXG acknowledges support from Jiangsu Funding Program for Excellent Postdoctoral Talent.

The LBT is an international collaboration of the University of Arizona, Italy (INAF: Istituto Nazionale di Astrofisica), Germany (LBTB: LBT Beteiligungsgesellschaft), The Ohio State University, representing also the University of Minnesota, the University of Virginia, and the University of Notre Dame.
The WFST team would like to express their sincere thanks to the staff of the Lenghu observation station.

This work has made use of data from the European Space Agency (ESA) mission {\it Gaia} (\url{https://www.cosmos.esa.int/gaia}), processed by the {\it Gaia} Data Processing and Analysis Consortium (DPAC, \url{https://www.cosmos.esa.int/web/gaia/dpac/consortium}).
This work made use of Astropy:\footnote{http://www.astropy.org} a community-developed core Python package and an ecosystem of tools and resources for astronomy \citep{astropy:2013, astropy:2018, astropy:2022}.
}

\bibliography{main.bib}

\end{document}